# Structural relaxation around substitutional $Cr^{3+}$ in pyrope garnet

Amélie Juhin,[1,*] Georges Calas,[1] Delphine Cabaret,[1] Laurence Galoisy,[1] and Jean-Louis Hazemann[2]

[1]Institut de Minéralogie et Physique des Milieux Condensés (IMPMC), UMR CNRS 7590 Universités Paris VI et VII, IPGP, 4 Place Jussieu, 75252 Paris Cedex 05, France

[2]Institut Néel, CNRS - Université Joseph Fourier, 25 avenue des Martyrs, BP 166, 38042 Grenoble cedex 9, France

**ABSTRACT**

The structural environment of substitutional $Cr^{3+}$ ion in a natural pyrope $Mg_3Al_2Si_3O_{12}$ has been investigated by Cr K-edge Extended X-ray Absorption Fine Structure (EXAFS) and X-ray Absorption Near Edge Structure (XANES) coupled with first-principles computations. The Cr-O distance is close to that in knorringite $Mg_3Cr_2Si_3O_{12}$, indicating a full relaxation of the first neighbors. The local $C_{3i}$ symmetry of the octahedral Y site is retained during Cr-Al substitution. The second and third shells of neighbors (Mg and Si) relax only partially. Site relaxation is accommodated by strain-induced bond buckling, with angular tilts of the Si-centered tetrahedra around the Cr-centered octahedron, and by a radial deformation of the Mg-centered dodecahedra.

---

[*] E-mail : Amelie.Juhin@impmc.jussieu.fr

# INTRODUCTION

Cr-bearing pyrope is an important mineral of the lithospheric upper mantle. Chromium concentration in this mineral increases with depth and is used in mantle barometry (see e.g., Grütter et al. 2006). The presence of Cr in the garnet lattice shifts significantly the spinel-garnet transformation -a major phase boundary in the Earth's upper mantle- to pressures as high as 7 GPa (Klemme, 2004). Pyrope $Mg_3Al_2Si_3O_{12}$ and knorringite $Mg_3Cr_2Si_3O_{12}$ form a complete solid solution and garnets close to the pyrope 50-knorringite 50 composition occur as inclusions in natural diamonds (Irifune et al. 1982). In the garnet structure (Novak and Gibbs 1971), $Cr^{3+}$ is small enough to substitute $Al^{3+}$ in the Y site, at the center of a slightly distorted octahedron of $C_{3i}$ (or $\bar{3}$) symmetry, giving rise to characteristic optical absorption spectra (Amthauer 1976). The Cr-O distance inferred from crystal field splitting indicates a partial relaxation of the Cr-site in pyrope (Taran et al. 2004). Although $Sc^{3+}$ substitution has been recently investigated in garnets (Quartieri et al. 2006), direct structural information is still lacking for $Cr^{3+}$. In the $\alpha$-$Al_2O_3$ - $\alpha$-$Cr_2O_3$ and $MgAl_2O_4$ - $MgCr_2O_4$ solid solutions, the Cr to Al substitution induces a significant relaxation, with an extension depending on the host structure (Gaudry et al. 2006; Juhin et al. 2007).

The present work investigates the relaxation caused by the substitution of $Al^{3+}$ by $Cr^{3+}$ in pyrope. The combination of X-ray Absorption spectroscopy and Density Functional Theory (DFT) calculations demonstrates a full relaxation of the substituted Y site, the structural relaxation extending at least to the third neighbor shell.

# EXPERIMENTAL DETAILS

We investigated a natural gem-quality pyrope single crystal from Garnet Ridge, Arizona, brought up by ultramafic diatremes of the Navajo Volcanic Field and considered as characteristic of the underlying shallow upper mantle (Wang et al. 1999). The pyrope contains 40.6 wt%-$SiO_2$, 19.8 wt%-$MgO$, 22.3 wt%-$Al_2O_3$, 2.09 wt%-$Cr_2O_3$, 9.59 wt%-$FeO$ and 4.01 wt%-$CaO$. The composition was analyzed using the Cameca SX50 electron microprobe at the CAMPARIS facility (Universities of Paris 6/7, France). A 15 kV voltage with a 40 nA beam current was used. X-ray intensities were corrected for dead-time, background, and matrix effects. The standards used were $CaMgSi_2O_6$, $KAlSi_3O_8$, $\alpha$-$Fe_2O_3$ and $\alpha$-$Cr_2O_3$.

Cr K-edge X-ray Absorption spectra were collected at room temperature on beamline BM30b (FAME), at the European Synchrotron Radiation Facility (Grenoble, France). Beam conditions and details of experiment are given in Juhin et al. (2007). Data treatment and EXAFS analysis were performed using the IFEFFIT program suite (Ravel and Newville 2005). A multiple-shell fit was performed in the R-space between 1.1 and 3.5 Å within the k-range 3.7-10.2 Å$^{-1}$, including the first four single scattering paths (O, Mg, Si and O). Mg and Si paths were treated identically. We used a single amplitude parameter for all paths, one single energy shift $\Delta e_0$ (fitted to -0.1 eV) for all paths, two distinct mean-square displacement values $\sigma^2$ for the oxygen shells and the non–oxygen (fitted to 0.0030 Å$^2$ and 0.011 Å$^2$), and three different path lengths $\Delta R_i$. The R-factor was 0.017. A simulation of the structural relaxation was performed to quantify the geometric surrounding around an isolated $Cr^{3+}$. The first-principles calculations (based on DFT-LSDA) were done in a neutral unit cell of $Mg_3Al_2Si_3O_{12}$ (160 atoms), with plane wave basis set and norm-conserving pseudopotentials. For Si, we considered 3s, 3p, 3d as valence states (core radii of 1.05 a.u, $\ell$=2 taken as local part) and those of Juhin et al. (2007) for Mg, Al, Cr, O. This unit cell was first relaxed with cut-off energy of 70 Ry and sampling of the Brillouin Zone (BZ) at $\Gamma$+ (1/2, 1/2, 1/2). We obtained a lattice constant of 11.26 Å (-1.6 % relative to experiment) and a general position of (0.0324, 0.0518, 0.6529) in good agreement with the experimental structure (Pavese et al. 1995). Then, an Al atom was substituted by a Cr in its exact position, building the "non-relaxed model". The "relaxed model" was obtained by relaxing all atomic positions with the same cut-off energy and BZ sampling, while the size of the supercell was kept fixed. The minimal Cr-Cr distance (11.26 Å) is large enough to minimize interactions between two impurities. To compare directly the theoretical bond distances to the experimental ones, we rescaled the lattice parameter by 1.6 %. This operation does not affect the reduced atomic positions. Additionally, the XANES spectrum was computed in the electric dipole approximation for both the relaxed and non-relaxed models. The method used for XANES calculations is described in Taillefumier et al. (2002). We used the same code and pseudopotentials, except that of Cr, which was generated with only one 1s electron. Convergence was reached for a 70 Ry energy cut-off for the plane-wave expansion, one k-point for the charge density calculation, and a Monkhorst-Pack grid of $2\times2\times2$ k-points in the BZ for the absorption cross-section calculation. The continued fraction is computed with a constant broadening $\gamma =1.1$ eV, which takes into account the core-hole lifetime.

**RESULTS AND DISCUSSION**

**Evidence of a full relaxation at the Cr-site in pyrope**

The Fourier Transform shows the contribution of the various combination shells around substitutional $Cr^{3+}$ (Figure 1). Calculated and experimental Cr-O distances are equal to 1.96 Å (± 0.01 Å), with six oxygen first neighbors (Table I). The low Debye-Waller factor (0.055 Å) indicates the absence of radial distortion of the Y site. The Cr-O distance is similar to that in knorringite (1.96 Å) but larger than the Al-O distance in pyrope (1.89 Å), which may be related to the difference in ionic radii, between octahedral $Cr^{3+}$ and $Al^{3+}$, 0.615 Å and 0.535 Å, respectively (Shannon and Prewitt, 1969). This demonstrates the existence of a full structural relaxation around $Cr^{3+}$. Using the relaxation parameter of Martins and Zunger (1984), here defined as:

$$\varsigma = \frac{R_{Cr-O}(Mg_3Al_2Si_3O_{12}:Cr^{3+}) - R_{Al-O}(Mg_3Al_2Si_3O_{12})}{R_{Cr-O}(Mg_3Cr_2Si_3O_{12}) - R_{Al-O}(Mg_3Al_2Si_3O_{12})}, \quad (1)$$

this corresponds to the relaxation limit $\zeta = 1$, as in $MgAl_2O_4$ spinel (Juhin et al. 2007). This value is higher than the one obtained by deriving the mean Cr-O distance from optical absorption spectra in a point charge model approach ($\zeta = 0.77$, Taran et al. 2004). This is also larger than $\zeta = 0.76$ in ruby (Gaudry et al. 2006). For $Sc^{3+}$ entering the Y site of andradite, only a partial relaxation was observed, with $\zeta = 0.47$ (calculated from Quartieri et al. 2006, for an andradite sample with 2.71 wt%-$Sc_2O_3$). These observations confirm that relaxation processes are a common feature of impurity insertion in mineral lattices, demonstrating the strong limitation of the Vegard law in solid solutions (Galoisy 1996). The evidence of relaxation provides also support for the elastic-strain theory, used to rationalize the incorporation of elements in mineral structures (Allan et al. 2003).

The symmetry of the relaxed Cr-site in Cr-bearing pyrope is similar to that of the Al-site in pyrope and to that of the Cr-site in knorringite. It belongs to the $C_{3i}$ point group, with an inversion center and a ternary axis, which is perpendicular to the plane formed by the $O^5$, $O^{11}$ and $O^{12}$ atoms (Figure 2). The presence of an inversion center is consistent with the low value of the molar extinction coefficient of $Cr^{3+}$ in pyrope (Taran et al. 1994). The preservation of the trigonal distortion after substitution at the Y site has also been observed with Electron Paramagnetic Resonance in $Cr^{3+}$-bearing Yttrium-aluminium and Yttrium-gallium garnets (Carson and White 1961). The Cr-centered octahedron is slightly more distorted in $Cr^{3+}$- bearing pyrope, with O-Cr-O angles of 86.3°, than in unsubstituted pyrope (O-Al-O angles of 87.3°) and in knorringite (O-Cr-O angles of 88.0°) (Novak and Gibbs, 1971). Both shared and unshared edges of the octahedron lengthen.

**Extension of relaxation around substitutional $Cr^{3+}$**

The theoretical Cr-Mg and Cr-Si distances in Cr-bearing pyrope, 3.22 Å and 3.23 Å, respectively, are intermediate between the distances, which separate octahedral Y sites from dodecahedral X and tetrahedral Z sites in pyrope and knorringite. This indicates that the Mg/Si second and third neighbors relax partially, as observed in spinel, whereas they are located at a larger distance from Cr. Indeed, because of the existence of deformable dodecahedral X sites, the bulk modulus B of pyrope is equal to 180 GPa (Pavese et al. 1995), a value smaller than that of spinel (200 GPa, Anderson and Nafe 1965). The calculated Cr-O and Al-O distances (3.55 Å) are equal, showing that the fourth neighbors are not affected by the substitution.

The reliability of this structural relaxed model was evaluated by a simulation of the Cr K-edge XANES of Cr-pyrope. All XANES features, labeled a to g, are reproduced by the simulation (Figure 3). Pre-edge features cannot be simulated in the electric dipole approximation since the Cr-site is centrosymmetric, as they arise from pure electric quadrupole transitions. The effect of the relaxation was evaluated by computing the XANES spectrum for the non-relaxed model. The main edge region (peaks b, c, d) is not correctly reproduced. Although the intensity, shape and position of features e and f are similar for the relaxed and non-relaxed models, the shape and relative intensity of peak g is improved in the relaxed model. The relaxed model provides the best agreement with experimental data.

The Cr-Mg and Cr-Si distances (3.32 Å) derived from EXAFS data are larger than the theoretical ones (Table I), even within the experimental uncertainty (±0.05 Å). Unexpectedly, they are also larger than those in knorringite, but the structure of this end-member has been only extrapolated but not refined. A possible explanation lies in the relative weakness of the second peak on the FT (Figure 1), despite the significant number of scattering atoms (Mg, Si and further O) contributing to it. Hence, it couldn't be fitted separately from the first peak, because of the reduced number of independent parameters, which prevented non-ambiguous assignments. The choice of other X cations, such as Ca/Fe or mixed shell of Mg, Ca and Fe according to sample composition, did not improve the fit. Nevertheless, XANES simulation using the relaxed model is in favor of a second shell mostly composed of Mg atoms, consistently with the relative abundance of Mg, Ca and Fe atoms in the sample.

**The relaxation process in pyrope**

The full relaxation of the oxygen first neighbors is partially accommodated by the slight angular distortion (1°) of the central octahedron and the radial shift of the Mg and Si neighbors. Our theoretical work shows that the $MgO_8$ dodecahedra connected to the central $CrO_6$ octahedron undergo a deformation, which leads to the lengthening of the average X1-O distance and to the dispersion of the X1-O and X2-O distances. The tetrahedra rotate around the $\bar{4}$ axis. The position angle α, which describes the rotation around the $\bar{4}$ axis (Born and Zeeman 1964), varies from 28.0° to 30.1° for the $O^1$-$O^2$ edge but only to 28.3° for the $O^3$-$O^{13}$ edge: this indicates a slight distortion of the tetrahedra connected to $CrO_6$. Indeed, rigid unit modes are forbidden in the garnet structure, and relaxation can only be achieved by the distortion of the polyhedra and rotation of the more rigid tetrahedra. (Ungaretti et al. 1995).

The importance of structural relaxation during Cr to Al substitution in pyrope likely arises from the energetic aspects of the process. DFT simulations give a relaxation energy - defined as the energy difference between non-relaxed and relaxed models- of +41 kJ.mol$^{-1}$. Calculations favor the stability of the relaxed structure, showing the major importance of the energetics of the chemical bonds with the nearest neighbors. The positive effect of pressure on chromium concentration in pyrope may arise from the compression of Cr-bearing Y sites, which will reduce the local stress resulting from the Cr to Al substitution. The combined approach presented in this study presents new perspectives for linking structural, electronic and energetic aspects of the relaxation around impurities in minerals.

## ACKNOWLEDGMENTS

The theoretical part of this work was supported by the Institut du Développement et de Recherche en Informatique Scientifique under project 62015. This is IPGP Contribution n°XXXX.

Wang, L., Essene, E.J., and Zhang, Y. (1999) Mineral inclusions in pyrope crystals from Garnet Ridge, Arizona, USA: implications for processes in the upper mantle. Contributions to Mineralogy and Petrology, 135, 164-178.

**Figures**

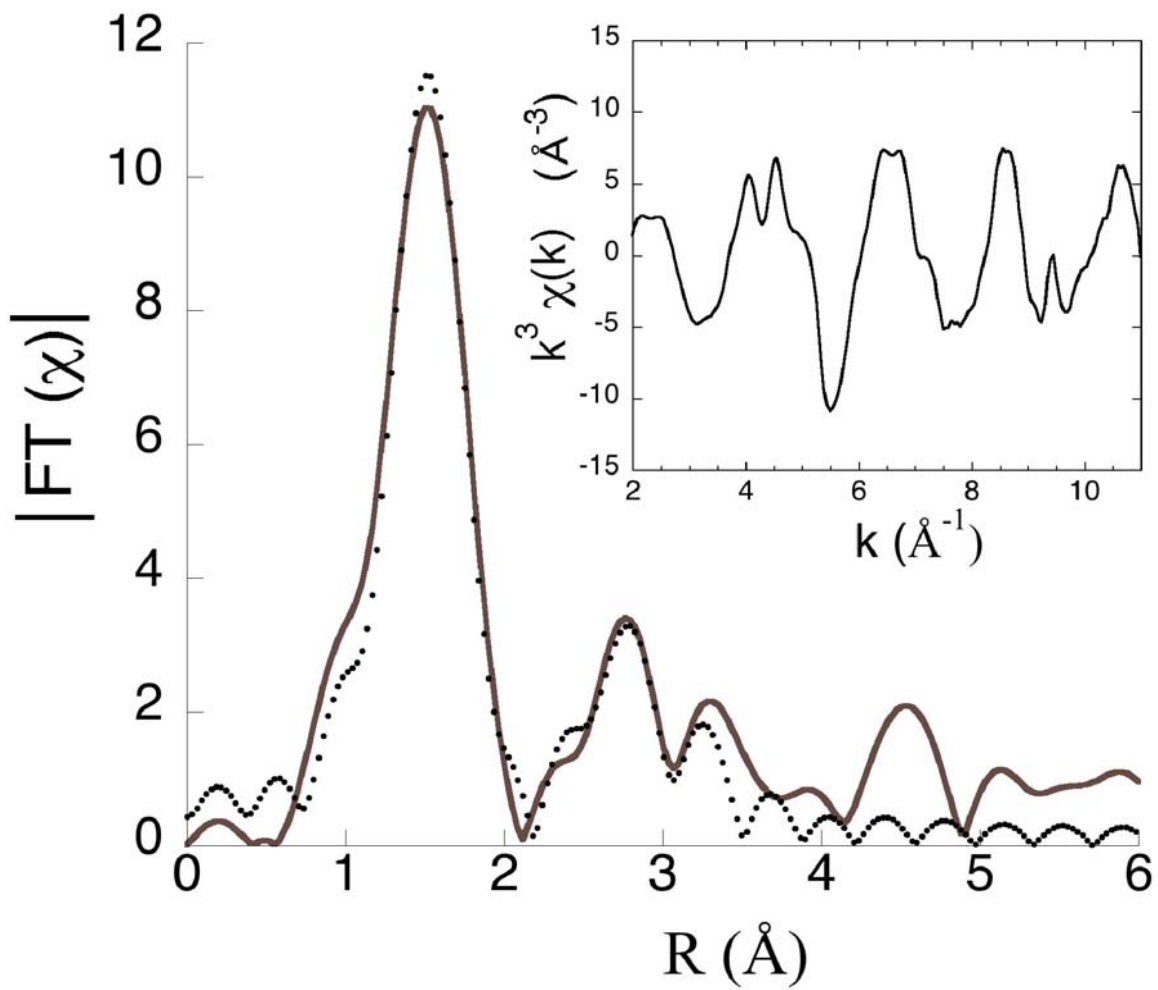

Figure 1. Fourier transform of the EXAFS data (solid line) and fit model (dots) for Cr-doped pyrope. The inset reports the $k^3$-weighted $\chi(k)$.

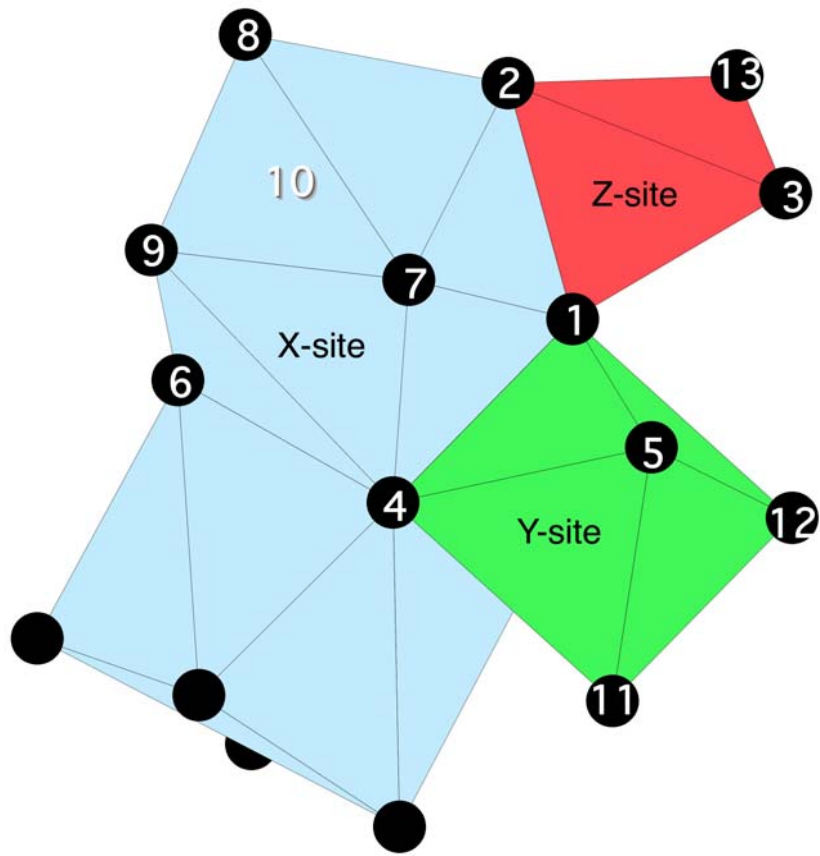

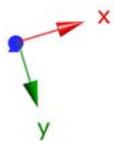

Figure 2. (Color online) Portion of the garnet structure. The oxygen atoms are labeled according to Novak and Gibbs (1971), with additional labels 9, 10, 11, 12 and 13. The trigonal axis is perpendicular to the plane formed by the $O^5$, $O^{11}$ and $O^{12}$ atoms, along the (111) direction. It is retained after substitution of Al for Cr, as well as the inversion.

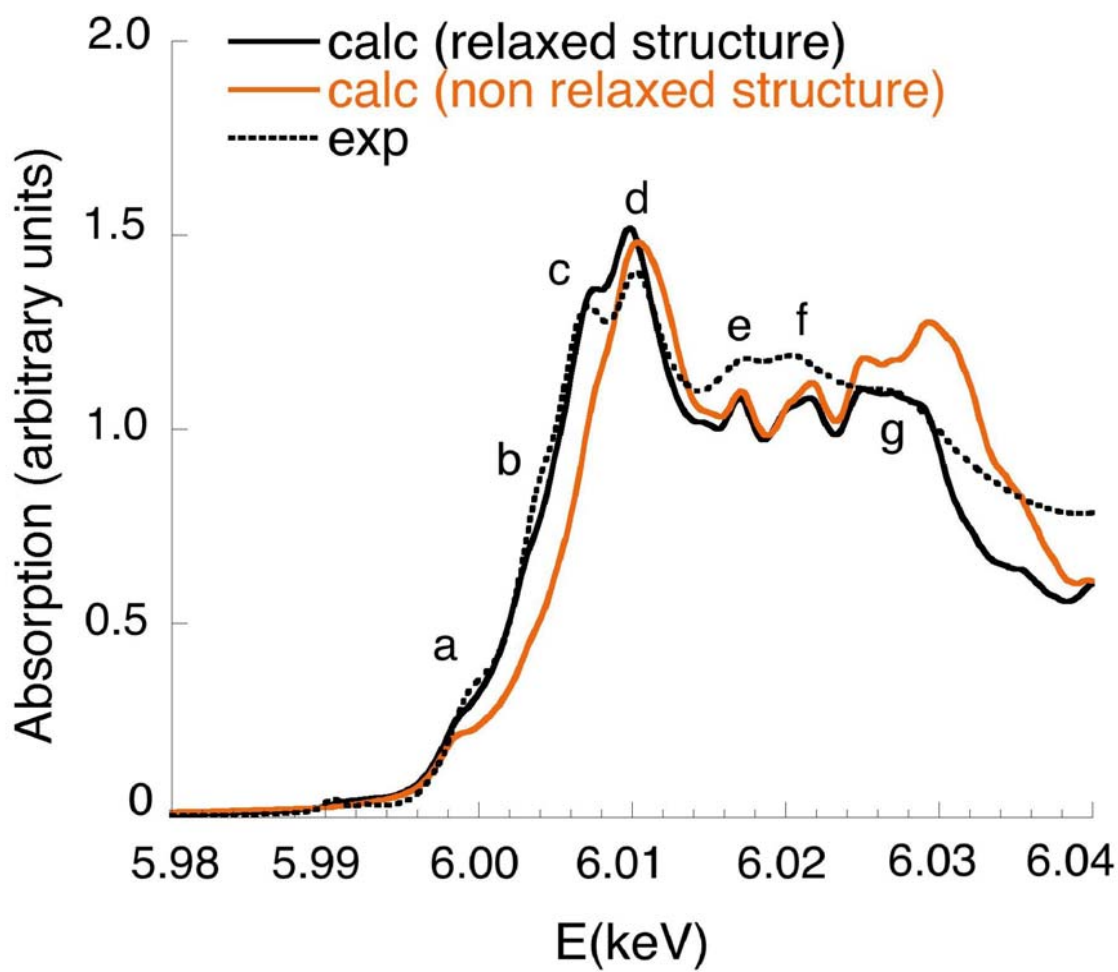

Figure 3. (color online) Cr K-edge XANES spectra in $Mg_3Al_2Si_3O_{12}$: $Cr^{3+}$. The experimental signal (dotted line) is compared with the theoretical spectra calculated in the relaxed model (black solid line) and in the non-relaxed model (orange solid line). Both calculated spectra have been normalized to the experimental one with the same factor. The pre-edge is visible at 5.991 keV.

**Table**

TABLE 1. Calculated and experimental distances (Å), angles inside and between polyhedra (degrees), for the different structures considered in this work. The oxygen's superscripts correspond to the Novak and Gibbs labeling (see Figure 2). The numbers in parenthesis indicate the multiplicity of the bond distances.

| | | $Mg_3Al_2Si_3O_{12}$ exp * | $Mg_3Al_2Si_3O_{12}$ calc | $Mg_3Al_2Si_3O_{12}:Cr^{3+}$ exp | $Mg_3Al_2Si_3O_{12}:Cr^{3+}$ calc | $Mg_3Cr_2Si_3O_{12}$ exp † |
|---|---|---|---|---|---|---|
| Y - O | Cr - O | --- | --- | (6) 1.96 ± 0.01 | (6) 1.96 | (6) 1.96 |
| | Al - O | (6) 1.887 | (6) 1.89 | --- | --- | --- |
| Y - X | Cr - Mg | --- | --- | (6) 3.30 ± 0.05 | (6) 3.22 | (6) 3.25 |
| | Al - Mg | (6) 3.202 | (6) 3.20 | --- | --- | --- |
| Y - Z | Cr – Si | --- | --- | (6) 3.30 ± 0.05 | (6) 3.23 | (6) 3.25 |
| | Al – Si | (6) 3.202 | (6) 3.20 | --- | --- | --- |
| Y - O | Cr - O | --- | --- | --- | (6) 3.55 | (6) 3.57 |
| | Al - O | (6) 3.555 | (6) 3.55 | --- | --- | --- |
| Z – O | Si – O | (4) 1.634 | (4) 1.65 | --- | (4) 1.65 | (4) 1.64 |
| X1 – O | Mg – O $^{9,1,6,2}$ | (4) 2.198 | (4) 2.20 | --- | (3) 2.20, 2.22 | (4) 2.24 |
| X2 – O | Mg – O $^{4,8,10,7}$ | (4) 2.341 | (4) 2.32 | --- | 2.29, 2.30, 2.33, 2.34 | (4) 2.36 |
| X-edges | $O^4 – O^{6,7}$ | 2.708, 2.777 | 2.69, 2.76 | --- | 2.64, 2.74 | 2.70, 2.78 |
| Y-edges | $O^1 – O^{4,5}$ | 2.618, 2.73 | 2.60, 2.73 | --- | 2.69, 2.87 | 2.72, 2.82 |
| Z-edges | $O^1 – O^{2,3}$ | 2.496, 2.750 | 2.52, 2.77 | --- | 2.51, 2.79 | 2.51, 2.76 |
| $O – \hat{Y} - O$ | $O^1 – Cr – O^4$ | --- | --- | --- | 86.3 | 87.8 |
| | $O^1 – Al – O^4$ | 87.88 | 87.3 | --- | --- | --- |
| Alpha | Edge $O^1 - O^2$ | 27.50 | 28.0 | --- | 30.1 | 29.4 |
| | Edge $O^3 - O^{13}$ | 27.50 | 28.0 | --- | 28.3 | 29.4 |

Notes: * Pavese et al. 1995; † Novak and Gibbs. 1971